\documentclass[runningheads]{llncs}

\usepackage{graphicx}
\usepackage{tabularx}
\usepackage{perpage}
\MakePerPage{footnote}
\usepackage{hyperref}

\begin{document}

\title{High-resolution imaging on TPUs \thanks{This work was conducted at Google.}
}

\author{Fantine Huot\inst{1} \and
Yi-Fan Chen\inst{2} \and
Robert Clapp \inst{1} \and
Carlos Boneti\inst{2} \and
John Anderson\inst{2}}

\authorrunning{F. Huot et al.}

\institute{Stanford University, Stanford CA 94305, USA \\
\email {\{fantine,bob\}@sep.stanford.edu} \and
Google, Mountain View CA 94043, USA \\
\email{\{yifanchen,carlosboneti,janders\}@google.com}}

\maketitle  
\begin{abstract}
The rapid evolution of artificial intelligence (AI) is leading to a new generation of hardware accelerators optimized for deep learning. Some of the designs of these accelerators are general enough to allow their use for other computationally intensive tasks beyond AI. Cloud tensor processing units (TPUs) are one such example. Here, we demonstrate a novel approach using TensorFlow on Cloud TPUs to implement a high-resolution imaging technique called full-waveform inversion. Higher-order numerical stencils leverage the efficient matrix multiplication offered by the Cloud TPU, and the halo exchange benefits from the dedicated high-speed interchip connection. The performance is competitive when compared with Tesla V100 graphics processing units and shows promise for future computation- and memory-intensive imaging applications. 

\keywords{Imaging \and TPU \and Full-waveform inversion.
}
\end{abstract}

\section{Introduction}
Full-waveform inversion (FWI) is a computationally intensive, high-resolution imaging algorithm used to image volumes from waveform data acquired at the surface, such as for geophysical imaging or medical imaging. FWI is an iterative data-fitting procedure based on full-wavefield modeling and is used to extract quantitative information from recorded wavefields~\cite{fichtner2008theoretical,virieux2009overview}. FWI extends refraction- and reflection-tomography techniques~\cite{clement2001migration,plessix1998waveform}, which only use the travel-time kinematics of the wave, by  means of additional information provided by the amplitude and phase of the waveform. Given an initial guess of the parameters of the imaged volume, wavefield data are predicted by solving a wave equation, followed by updating the volume parameters in order to reduce the misfit between observed and predicted data. This step is repeated iteratively until the data-misfit is sufficiently small. Misfit evaluation and computation of its gradient are expensive operations that involve solving numerical partial differential equations for each iteration. As a consequence, a key requirement of FWI is an efficient wavefield-modeling engine. 

Advances in high-performance computing have made FWI computationally feasible,
and it has become one of the state-of-the-art methods used for Earth subsurface imaging. FWI has led to significant breakthroughs in imaging complex geology and identifying potential geohazards~\cite{ben2008velocity,brossier2009seismic,sirgue2004efficient,vigh20083d}. Recently, FWI demonstrated the potential to improve medical imaging with ultrasound~\cite{bernard2017ultrasonic,calderon20173d,lin2012ultrasound,sandhu20173d} as a safe, non-invasive procedure that does not use radiation. Ultrasound medical imaging can (i) help diagnose heart conditions, or assess damage after a heart attack, (ii) diagnose causes of pain, swelling and infection, and (iii) examine fetuses in pregnant women or the brain and hips in infants. Ultrasound medical imaging is traditionally performed using reflection tomography; however, FWI shows promise for resolving more complex waveforms, potentially leading to new applications, including breast cancer detection~\cite{calderon20173d,lin2012ultrasound,sandhu20173d}.

A primary drawback of FWI is its greater computational cost. Methodologies to reduce this computational cost, especially for three-dimensional (3D) multicomponent data, are the subject of ongoing research \cite{fichtner2013multiscale,metivier2013full,vigh20083d}. FWI requires a large amount of high-performance computing resources that are usually available in multicore computing architectures based on distributed shared memory or distributed clusters with homogeneous or heterogeneous nodes commonly seen in private or commercial clouds. As a benefit of the rapid evolution of artificial intelligence (AI), commercial clouds now provide not only central processing units (CPUs) and graphics processing units (GPUs) but also new hardware accelerators optimized for deep learning. The Cloud tensor processing unit (Cloud TPU) is an AI application-specific integrated circuit  developed by Google for neural-network machine learning and has received much attention in the machine-learning community~\cite{jouppi2017datacenter,kurian2019domain}. Its latest release, Cloud TPU version 3, comprises four chips, can achieve 420 tera-FLOPS (floating-point operations per second), and has 128GB of high-bandwidth memory (HBM)\footnote{cloud.google.com/tpu}. Multiple units are connected to form a “Pod” (Cloud TPU v3 Pod) through a dedicated high-speed torus network, allowing up to 100+ peta-FLOPS and 32TB of HBM to be accessed by the application with very low latency. TPUs are programmable via software frontends, such as TensorFlow~\cite{abadi2016tensorflow} or PyTorch~\cite{paszke2017pytorch}, and can be deployed to accelerate both training of deep neural networks and online prediction~\cite{jouppi2017quantifying}.

With the tremendous amount of computational resources that TPUs offer, it is compelling to also consider the opportunities TPUs bring for applications beyond machine learning. The programming frontends that are used for TPUs, such as TensorFlow, offer a rich set of functionalities that are highly relevant for scientific computing. The TensorFlow TPU programming stack provides the additional benefit of allowing distributed algorithms to be expressed with simple code without sacrificing performance. A previous study \cite{yang2019high} presents a compelling example of using the TensorFlow TPU framework for scientific computing. 

On the basis of these observations, we investigated leveraging Cloud TPUs for FWI by adapting numerical schemes to Cloud TPUs and demonstrating the suitability of TPUs for wavefield modeling. These optimizations resulted in our implementation prototype showing  competitive performance relative to Tesla V100 GPUs and demonstrated our approach as promising for future waveform-inversion applications. 

In the following sections, we review the architecture of the TPU framework and discuss the adaptation of the wavefield-modeling engine to better leverage the TPU architecture. We then present preliminary benchmarks and discuss the results. 
\section{TPU Device Architecture}
TPUs are programmable linear algebra accelerators optimized for machine-learning workloads. In the Cloud TPU version 3 architecture, one “TPU unit” comprises four TPU chips on a board, with each TPU chip containing two TensorCores, which are treated as independent processors that communicate with each other through a dedicated high-bandwidth, low-latency, interchip network. In a larger system, $>$1000 TPU chips are packed onto a two-dimensional (2D) toroidal mesh interchip interconnected network to form a TPU cluster (“TPU Pod”)~\cite{yang2019high}. Each TPU unit is paired with a TPU host server that includes a CPU, memory, and disk resources. The host server is typically used for data preparation and I/O tasks, whereas  the TPUs are leveraged for computationally intensive tasks. The dedicated high-speed mesh network allows all TensorCores in a Pod to work in lockstep efficiently without accessing the host servers. 

\begin{figure}
\center
\includegraphics[width=0.9\textwidth]{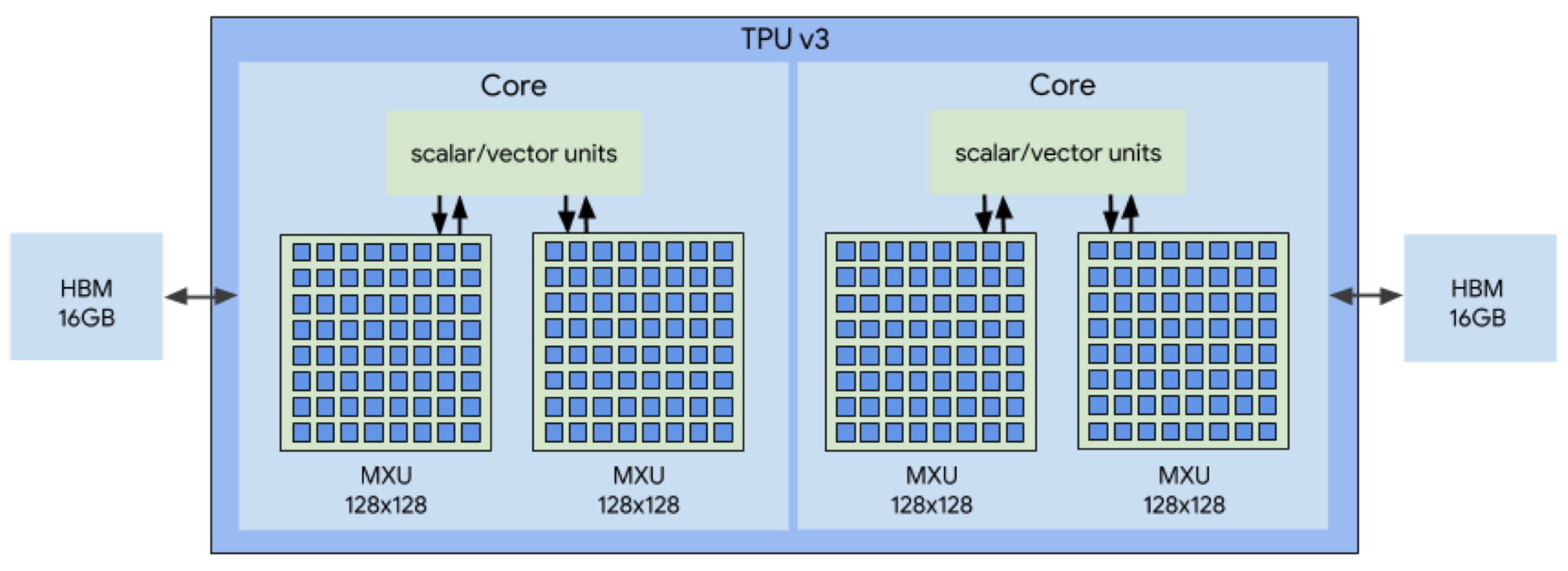}
\caption{One TPU chip comprises two TensorCores. A TensorCore of third-generation TPUs (TPU v3) includes a scalar processor, a vector processor, and two matrix units. The arrows depict the connections across different units and HBM. This diagram is borrowed from the system architecture guide~\cite{google2019system}, where “Core” corresponds to the TensorCore.} \label{tensorcore}
\end{figure}

The TensorCore, depicted in Figure~\ref{tensorcore}, is optimized for dense linear algebra computations and contains different classes of computing units, such as a scalar processor, a vector processor, accumulators, and matrix units~\cite{google2019system,jouppi2017datacenter,kurian2019domain}, all backed by 16GB of HBM. Vectorized operations are either handled directly by the vector processor or forwarded to the corresponding extended vector units, each of which accepts the input operands, performs the corresponding operations, and returns the results to the vector processor. One of these extended vector units is the matrix unit (MXU), which performs 128 $\times$ 128 multiply accumulate operations at each cycle [8]. The MXU represents the main computing power of the TPU architecture and should be exploited as much as possible. Although its inputs and outputs are 32-bit floating-point values, the MXU rounds inputs down to bfloat16, a 16-bit floating-point representation that provides better training and model accuracy than the IEEE half-precision representation, before multiplying.

\begin{figure}
\center
\includegraphics[width=0.5\textwidth]{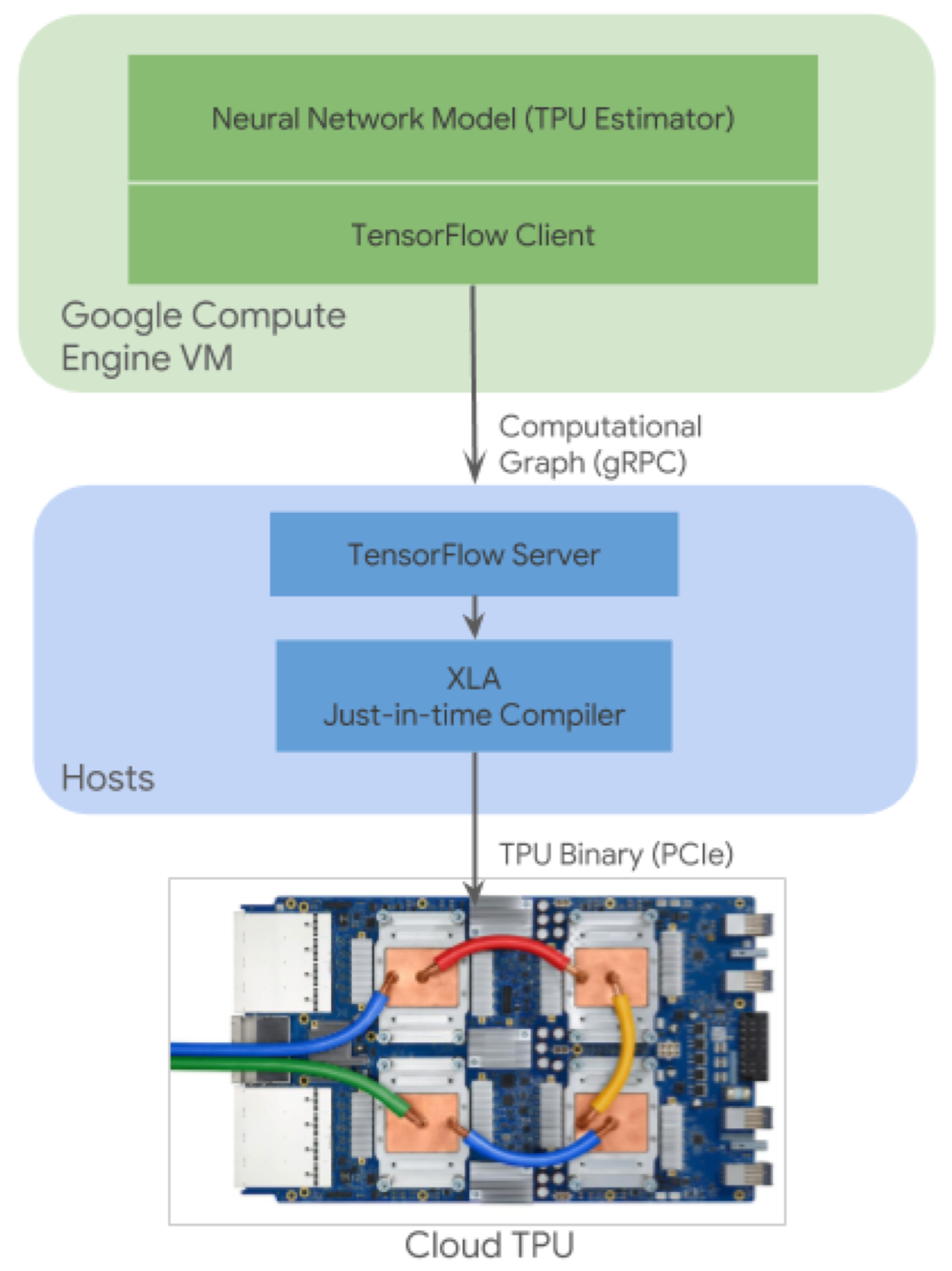}
\caption{The TPU software architecture comprises the neural-network model or other computational tasks, the TensorFlow client, the TensorFlow server, and the XLA compiler~\cite{google2019system}.} \label{xla}
\end{figure}

We program the TPUs through TensorFlow, with the flow used to run programs on TPUs using the TensorFlow frontend schematically depicted in Figure~\ref{xla}. Additional details are available on the official TensorFlow XLA page\footnote{www.tensorFlow.org/xla}. In the first stage, TensorFlow constructs the computation graph and marks it for replication. The graph is then rewritten to be TPU compatible and compiled to a high-level optimizer (HLO) program. Next, the accelerated linear algebra (XLA) compiler converts the HLO operations to low-level optimizer (LLO) code, effectively “TPU assembly code”, which can be readily executed on the TPUs. Graph construction and compilation occur on the host server and incur overhead cost; however, once the compiled LLO code is deployed to the TPUs, the computation step can be repeated as many times as necesssary without the intervention from the host servers. Additionally, XLA provides specific communication primitives implemented over the dedicated high-bandwidth, low-latency, interchip interconnected network. As a consequence, communication between the TPU chips within a TPU Pod is extremely efficient.

The shape of the tensor variables used in the program (expressed as TensorFlow tensors) can critically affect performance and memory usage. According to the performance guide~\cite{google2019performance}, unlike most other architectures, arrays in TPUs are tiled in two dimensions. Therefore, the program must be designed to enable XLA to perform the correct data-layout transformations by arranging the data in memory to allow their efficient processing by the hardware. Programs that operate on array sizes undividable by eight will exhibit suboptimal performance; therefore, dimensions should be padded to multiples of eight. 

\section{Wavefield-Modeling Engine}
FWI computes the parameters of the imaged volume, including propagation velocity, $v$, by minimizing the misfit between the modeled data, $d_{mod}$, and the observed data, $d_{obs}$. When using the least-squares misfit, the objective function, $\phi$, can be expressed as follows: 
\begin{equation}
\min_{v} \phi = ||d_{mod} - d_{obs} ||^2_2, 
\end{equation}
where $||.||_2$ represents the L2 norm and $d_{mod}$ is subject to the wave equation. 

The objective function is solved iteratively by computing the waveforms at each iteration. Therefore, an efficient wavefield-modeling engine is key. Here, we focus on the acoustic wave equation in a 3D isotropic constant-density medium: 
\begin{equation}
\frac{\partial^2 P(x,y,z,t)}{\partial t^2} = v(x,y,z)^2 \nabla^2 P(x,y,z,t) + s(x,y,z,t),
\end{equation}
where $P$ represents the pressure field at a given location $(x,y,z)$ at a time $t$, $v$ is the propagation velocity at a given location, $s$ is the source injection at a given location at a certain time, and $\nabla^2$ is the Laplacian operator in space. Although we demonstrate the feasibility of TPU implementation for the acoustic isotropic case,  the TPU adaptations presented here still hold for variants of the wave equation, such as elastic propagation or anisotropy.

We implement the wave equation using finite difference in the time domain and use the second-order approximation of the second derivative in time. In discrete space, we transcribe the numerical scheme for wave propagation as follows: 

\begin{equation}
P_{ijk}^{n+1} =  - P_{ijk}^{n-1} + 2P_{ijk}^{n} + v_{ijk}^2 \triangle t^2 \nabla^2 P_{ijk}^{n} + \triangle t^2 s_{ijk}^n,
\end{equation}
where $ijk$ represents indices of the spatial coordinates, $n$, $n+1$, and $n-1$ correspond to the current, next, and previous time steps, respectively, and $\triangle t$ corresponds to the time delay between two consecutive steps. The source injection is implemented using a queued infeed. 

\begin{figure}
\center
\includegraphics[width=0.7\textwidth]{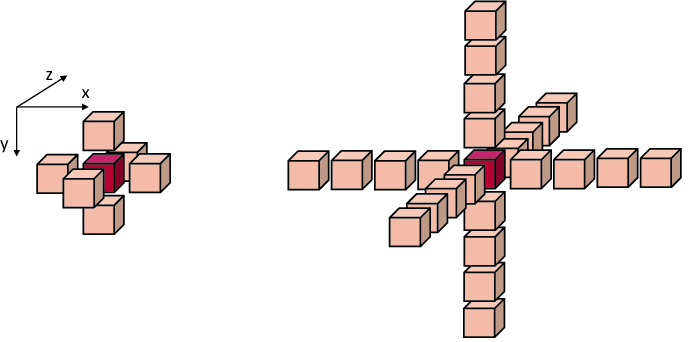}
\caption{Schematic representation of numerical stencils for the second-order (left) and eighth-order (right) approximation of the Laplacian in three dimensions. Here, we use numerical stencils of order $\geq$8.} \label{stencils}
\end{figure}

\begin{figure}
\center
\includegraphics[width=\textwidth]{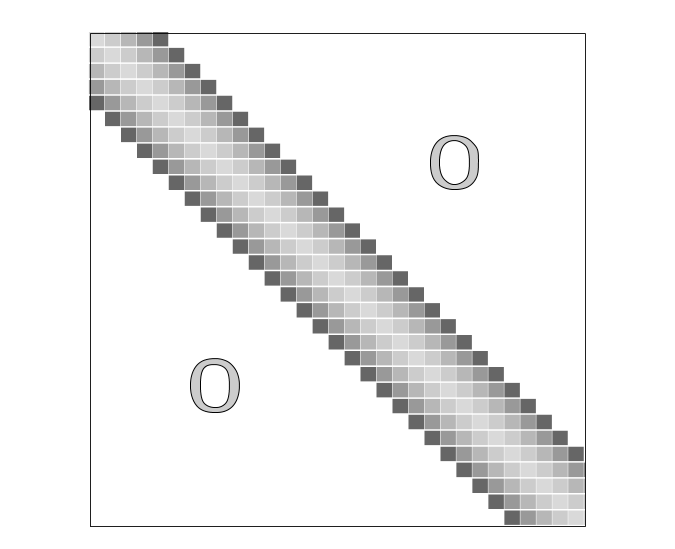}
\caption{The finite difference operations along the $x$ and $y$ axes are implemented as matrix multiplications with a constant band-diagonal operator to leverage the Cloud TPU MXU. Note the dimensions of the matrix are actually multiples of 128, following the performance guide \cite{google2019performance}.} \label{matmul-operators}
\end{figure}

Higher-order approximations of the Laplacian in space are necessary to avoid numerical dispersion. Figure~\ref{stencils} provides a schematic representation of the 3D numerical stencil. Here, we use numerical stencils of order $\geq$ 8. Within the TPU framework, we implemented the 3D pressure and velocity fields as slices of 2D TensorFlow tensors in order to explicitly conform to the recommended memory layout \cite{google2019performance}. To adapt to the TPU architecture, the numerical scheme for the Laplacian operator is expressed in multiple passes:  the finite-difference operators along the $x$ and $y$ axes are implemented as matrix multiplications or TensorFlow convolutions to leverage the TPU MXU, whereas the finite-difference operator along the $z$ axis is implemented using an element-wise vector operation (VPU). Figure~\ref{matmul-operators} represents the constant band-diagonal matrix for an eighth-order stencil. 

\begin{figure}
\center
\includegraphics[width=\textwidth]{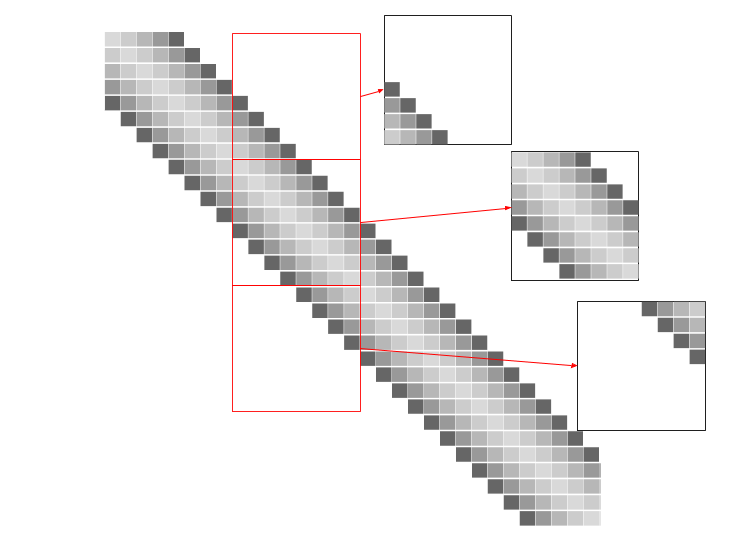}
\includegraphics[width=\textwidth]{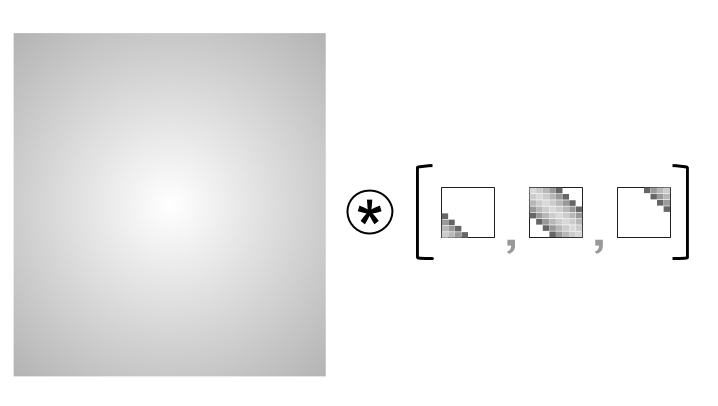}
\caption{Since the matrix multiplication operation on the $x$ and $y$ axes involves many MXU passes with zero 128 $\times$ 128 blocks due to the band-diagonal nature of the finite-difference matrix operator, the same operation can be expressed as a 1D TensorFlow convolution, allowing better MXU utilization.} \label{convol}
\end{figure}

Since the TPU MXU operates on 128 $\times$ 128 matrices~\cite{google2019performance}, finite difference operations on the $x$ and $y$ axes using matrix multiplication involve  subdividing the tensors into 128 $\times$ 128 blocks and performing multiple MXU passes. Due to the band-diagonal nature of the matrix operator (Figure~\ref{matmul-operators}), many of the resultant matrix multiplications involve zero blocks. Consequently, the operator can be written as a one-dimensional (1D) TensorFlow convolution, allowing performance gains while leveraging the TPU MXU (Figure~\ref{convol}). However, in both cases, the finite-difference operator along the $z$ axis relies on the TPU VPU. 

\section{Benchmark}

 We evaluated both the 8th- and 16th-order approximations of the Laplacian operator. There is a trade-off between the computation-domain size and the order of the numerical stencil, as a higher-order numerical stencil reduces the error in the approximation of the derivatives, thereby allowing for a larger sampling rate in grid cells and reducing the overall size of the computation domain. However, their use is often unpractical, because they incur a higher computational cost and significant memory exchange overhead due to the larger halos. In the case of TPUs however, the high-order stencils allow a better TPU-utilization rate by increasing the density of the numerical kernels. Additionally, the large memory exchange benefits from the dedicated high-speed, low-latency, interchip mesh network. 

\begin{figure}
\center
\includegraphics[width=0.9\textwidth]{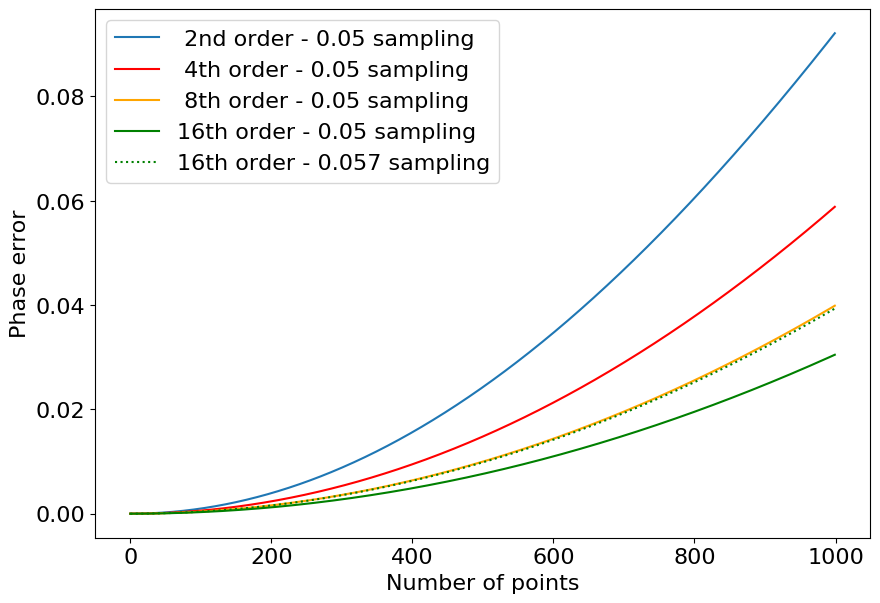}
\caption{Comparison of the phase error in the approximation of the second derivative of a function with a known second derivative (the cosine function) for varying stencil order and grid-cell sampling. The phase error decreases as the order of the stencil increases and subsequently increases along with larger grid-cell-sampling rates. Consequently, for the same tolerance in phase error, it is possible to reduce the computation domain by 12\% on each axis by transitioning from the 8th- to the 16th-order numerical stencil.} \label{stencil_order} 
\end{figure}

To evaluate the domain size and stencil trade-off, we compare the phase error in the approximation of the second derivative of a function with a known second derivative (the cosine function) (Figure~\ref{stencil_order}). The phase error decreases as the order of the stencil increases and subsequently increases along with larger grid-cell-sampling rates. Consequently, for the same tolerance in phase error, it is possible to reduce the computation domain by 12\% on each axis by transitioning from the 8th- to the 16th-order numerical stencil. When dealing with large 3D volumes, this domain reduction can represent significant savings in computational cost. 

\begin{table}
\center
\caption{Performance comparison between GPU baseline and TPU implementation with different stencil order and domain size.}\label{benchmark}
\begin{tabular*}{\textwidth}{c @{\extracolsep{\fill}} ccccc}
\hline
Architecture &  Stencil & Domain size & Number & Time (s) & Gcells/s \\
             &  order   & & of steps &  & \\
\hline
4 Tesla V100 GPU &  8th & 1024 $\times$ 1024 $\times$ 1024 & 100 & 8.98 & 11.96 \\
2$\times$2 TPU v3 &  8th & 1024 $\times$ 1024 $\times$ 1024  & 100 & 15.68 & 6.85 \\
2$\times$2 TPU v3 &  16th & 768 $\times$ 768 $\times$ 768 & 100 & 9.14 & 4.96 \\
\hline
\end{tabular*}
\end{table}

The benchmark results are presented in Table~\ref{benchmark}, both in total computation time and in number of gigacells  computed per second (Gcells/s). We compared our TPU implementation over  a 2 $\times$ 2 TPU v3 architecture with a GPU implementation \cite{leader2013large} involving four Tesla V100 GPUs. For optimal performance, the tensor sizes were multiples of 128, and wave propagation was benchmarked over 100 time steps following completion of the source injection. We observed that on a 1024 $\times$ 1024 $\times$ 1024 domain, our TPU implementation ran at about $\sim$50\% speed of the GPU implementation. By transitioning to the 16th-order stencil and increasing the grid spacing accordingly (to reduce the total domain size), the results of total computation time were comparable with the GPU benchmark. Although our TPU implementation is not as finalized as the GPU implementation, as we still expect improvements on the halo exchange, these initial results are promising. Additionally, our TPU implementation is written in easy-to-read high-level Python, whereas the GPU baseline is implemented in highly-optimized C++ and low-level CUDA, allowing researchers to easily adapt and improve our code for scientific computing. The large-scale, high-bandwidth, low-latency connectivity between TPU chips facilitates scaling the computation to larger problems. 

\section{Discussion} 
This study demonstrated the promise of using Cloud TPUs for high-resolution 3D imaging. We developed a novel approach to wavefield modeling using TensorFlow on Cloud TPUs by adapting a numerical scheme to exploit the TPU architecture, using higher-order stencils to leverage its efficient matrix operation and using the dedicated high-bandwidth, low-latency, interchip interconnected network of the TPU Pod for the halo exchange. Our preliminary benchmark demonstrated that given these adaptations, our TPU implementation is competitive with GPU implementations. Faster computation allows faster decision making in workflows involving high-resolution geophysical and medical imaging. Additionally, this work demonstrated how TensorFlow on Cloud TPU can be used to accelerate  conventional scientific simulation problems. Further development of TPU-implementation frameworks will potentially enable users to program TPUs with simple and easy-to-read code. 


\bibliographystyle{splncs04}
\bibliography{references}

\end{document}